\documentclass[aps, showpacs, showkeys, preprint, groupedaddress, preprintnumbers, amsmath, amssymb, floatfix]{revtex4}
\usepackage{graphicx}% Include figure files
\usepackage{dcolumn}% Align table columns on decimal point
\usepackage{bm}% bold math

\begin{document}

\title{Tuning Low Temperature Physical Properties of CeNiGe$_{3}$ by Magnetic Field.}
\author{E. D. Mun, S. L. Bud'ko, A. Kreyssig, P. C. Canfield}
\affiliation{Ames Laboratory US DOE and Department of Physics and Astronomy, Iowa State University, Ames, IA 50011, USA}%

\date{\today}

\begin{abstract}
We have studied the thermal, magnetic, and electrical properties of the ternary intermetallic system CeNiGe$_{3}$ by means of specific heat,
magnetization, and resistivity measurements. The specific heat data, together with the anisotropic magnetic susceptibility, was analyzed on the
basis of the point charge model of crystalline electric field. The $J$\,=\,5/2 multiplet of the Ce$^{3+}$ is split by the crystalline electric
field (CEF) into three Kramers doublets, where the second and third doublet are separated from the first (ground state) doublet by $\Delta_{1}$
$\sim$ 100\,K and $\Delta_{2}$ $\sim$ 170\,K, respectively. In zero field CeNiGe$_{3}$ exhibits an antiferromangeic order below $T_{N}$ = 5.0\,K.
For \textbf{H}\,$\parallel$\,\textbf{a} two metamagnetic transitions are clearly evidenced between 2\,$\sim$\,4\,K from the magnetization isotherm
and extended down to 0.4\,K from the magnetoresistance measurements. For \textbf{H}\,$\parallel$\,\textbf{a}, $T_{N}$ shifts to lower temperature
as magnetic field increases, and ultimately disappears at $H_{c}$ $\sim$ 32.5\,kOe. For $H\,>\,H_{c}$, the electrical resistivity shows the
quadratic temperature dependence ($\Delta\rho = A T^{2}$). For $H \gg H_{c}$, an unconventional $T^{n}$-dependence of $\Delta\rho$ with $n > 2$
emerges, the exponent $n$ becomes larger as magnetic field increases. Although the antiferromagnetic phase transition temperature in CeNiGe$_{3}$
can be continuously suppressed to zero, it provides an example of field tuning that does not match current simple models of Quantum criticality.
\end{abstract}

%%%%%%%%%%%%%%%%%%%%%%%%%%%%%%%%%%%%%%%%%%%%%%%%%%%%%%%%%%%%%%%%%%%%%%%%%%
%Use showkeys class option if keyword
                              %display desired
%%%%%%%%%%%%%%%%%%%%%%%%%%%%%%%%%%%%%%%%%%%%%%%%%%%%%%%%%%%%%%%%%%%%%%%%%%

\maketitle

\section{introduction}

The ground state of Ce-based intermetallic compounds is often determined by a competition between the intersite Ruderman-Kittel-Kasuya-Yosida
(RKKY) interaction and the onsite Kondo fluctuation, a competition that leads to a plethora of unusual ground states. Depending on the strength
of the hybridization between 4$f$ and conduction electrons relative to their coupling strength, the ground states broadly occur either as
magnetically ordered or nonmagnetic (paramagnetic). In addition to these two interactions, the crystalline electric field (CEF) plays a
significant role in determining the temperature-dependent thermodynamic and transport properties. With increasing hybridization, the magnetic
moments associated with localized 4$f$ electrons are increasingly screened by the conduction electrons, leading to a reduction of magnetic
moments \cite{Hewson}. The strength of the hybridization can be varied by suitable control parameters, such as pressure ($P$), composition ($x$),
and magnetic field ($H$). A quantum critical point (QCP) can be approached by tuning the second order phase transition temperature down toward
$T$\,$\rightarrow$ 0\,K with an external, non-thermal, control parameters ($P$, $x$, $H$) \cite{Stewart, Lohneysen}. Of these three parameters,
magnetic field is often considered to be an ideal control parameter, since it can be reversibly, and continuously, used to tune the system
towards the QCP. Many antiferromagnetic (AFM) heavy fermion (HF) compounds with field-induced QCP have been identified and exhibit prominent
non-Fermi liquid (nFL) behavior expanding out (to higher temperature) from the QCP, triggered by quantum fluctuations \cite{Gegenwart, Custers,
Balicas, Budko1, Niklowitz, Tokiwa}.

CeNiGe$_{3}$ has been studied and identified as an AFM Kondo system \cite{Pikul}. Strong anisotropy in thermodynamic and transport properties can
be expected from the crystal structure; CeNiGe$_{3}$ crystallizes in the orthorhombic SmNiGe$_{3}$-type structure ($Cmmm$, No 65) with lattice
parameters $a$ = 4.139\,\AA, $b$ = 21.828\,\AA, and $c$ = 4.1723\,\AA\,\cite{Pikul, Bodak}. A layered structure, CeNiGe$_{3}$ is assembled by a
sequence Ge$_{2}$Ni-CeGe$_{2}$Ce-NiGe$_{2}$ layers along the \textbf{b}-axis \cite{Pikul, Hyunjin}. In this structure, Ce occupies a single
4$j$-site with $mm2$ point symmetry. At ambient pressure and in zero field, CeNiGe$_{3}$ shows AFM order below $T_{N}$ = 5.5 K \cite{Pikul,
Nakashima, Kotegawa, Tateiwa}. Recent studies of polycrystalline CeNiGe$_{3}$ \cite{Pikul} have inferred that the AFM order can be suppressed by
external magnetic field, via metamagnetic transitions (MMT). Although polycrystalline samples can show intrinsic property of materials with
directional averaged value, single crystals are needed for reliable $H-T$ phase diagrams of magnetically anisotropic compounds. More recently,
pressure-induced superconductivity was found in electrical resistivity measurements on polycrystalline samples \cite{Nakashima, Kotegawa} and in
specific heat (ac calorimetry) measurements on single crystal samples \cite{Tateiwa}. These studies indicate that the RKKY and/or Kondo
interaction can be tuned by applied pressure and superconductivity can be induced in this Kondo system.

In order to gain insight into the anisotropic, low-temperature physical properties of CeNiGe$_{3}$, we have grown single crystals and performed
thermodynamic and transport measurements in applied magnetic fields up to 90\,kOe. In this paper specific heat, electrical resistivity, and
magnetization measurements on single crystal CeNiGe$_{3}$ as function of temperature and magnetic field are presented. A strong magnetic
anisotropy is observed and can be reproduced by CEF analysis, and two-step MMTs are clearly observed in the magnetization isotherms and
magnetoresistance data. The constructed $H-T$ phase diagram of CeNiGe$_{3}$, for applied magnetic field along the magnetically easy axis,
\textbf{H}\,$\parallel$\,\textbf{a}, is qualitatively in agreement with a previous polycrystalline sample study \cite{Pikul}, and is also similar
to the case of CeAuSb$_{2}$ \cite{Balicas} and the isostructural compound YbNiSi$_{3}$ \cite{Budko} in which the AFM order is also suppressed by
applied magnetic field via MMTs.

\section{Experimental}

Single crystals of $R$NiGe$_3$ ($R$ = Ce and Y) were grown via solution growth techniques from a Ni-Ge rich, self-flux using high purity starting
elements \cite{Fisk, Canfield1, Canfield2}. A Ni-Ge rich self-flux was used because of both the wide range of binary eutectic valley and the merit
of introducing no new elements in the growth. The constituent elements, using a starting molar proportion of 1\,:\,1.6\,:\,9 ($R$\,:\,Ni\,:\,Ge),
were placed in an alumina crucible and sealed in a silica tube under a partial pressure of Ar. The temperature of the furnace was raised to
1050\,$^{\texttt{o}}$C and after homogenizing the mixture for 2 hours, the melt solution was slowly cooled over 100 hours down to
780\,$^{\texttt{o}}$C for Ce, 850\,$^{\texttt{o}}$C for Y. After decanting the excess solution, using a centrifuge, thin, plate-like crystals
were obtained. Samples have shiny, and generally clean, surfaces; small, spot-like, flux droplets solidified on the crystal surface were polished
off before physical property measurements.

\begin{figure}
\centering
\includegraphics[width=0.6\linewidth]{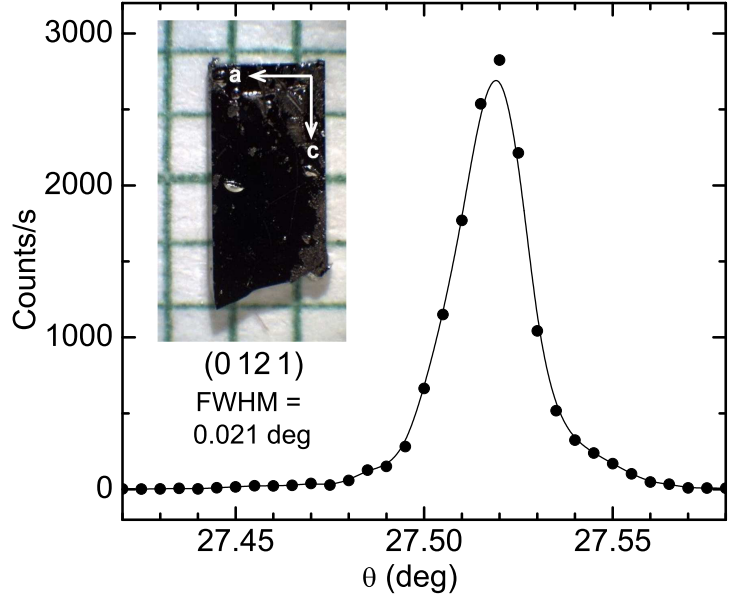}
\caption{X-ray rocking curve through the (0\,12\,1) reflection of the CeNiGe$_3$ single crystal. Inset shows a picture of a CeNiGe$_3$ single
crystal on mm grid paper. The orthorhombic \textbf{b} direction is perpendicular to the plate of the crystal. The short (longer) length of the
plate is parallel to the orthorhombic \textbf{a} (\textbf{c}) direction and the longer length of the plate is parallel to the $c$-axis. The small
droplets on the surface are residual flux that can be polished off before measurements.}
\label{xray}%
\end{figure}

The crystallographic \textbf{b}-axis was found to be perpendicular to the plate surface and well formed \textbf{a}-\textbf{c} edge facets were
clearly visible (Fig. \ref{xray}). The orthorhombic \textbf{b}-axis was easily determined to be perpendicular to the plate surface by x-ray Laue
pattern in reflection geometry. The orthorhombic \textbf{a} and \textbf{c} direction have been assigned by measurements of the lattice parameters
using a four-circle diffractometer with Cu-$K_{\alpha1}$ radiation from a rotating anode x-ray source, selected by a germanium (1\,1\,1)
monochromator. To confirm correct assignment, the reflection conditions according to the space group $C\,m\,m\,m$ has been satisfactory checked:
the (0\,12\,1) and (0\,14\,1) reflections have been found in contrast to the observation of pure background signal at what would be the (0\,13\,1)
position, in agreement with the general reflection condition $h$\,+\,$k$\,=\,2$n$\,=\,even for ($h$\,$k$\,$l$) Bragg reflections. The high quality
of the single crystals can be demonstrated by the excellent mosaicity of 0.021\,deg. in full width half maximum for the rocking scan shown in Fig.
\ref{xray}.

Powder x-ray diffraction measurements were taken on a Rigaku MiniFlex at room temperature with Cu-$K_{\alpha1}$ radiation in order to confirm the
crystal structure and lattice parameters. We prepared two powder samples of the $R$NiGe$_{3}$ ($R$ = Ce and Y) compounds, the first one was for
confirming the crystal structure and phase purity, the second one, which was mixed with Si-reference powder for correcting zero shift, was for
determining the lattice constants. No secondary phases were detected, except small amount of pure Ge phase coming from residual flux droplets on
the crystal surface. After polishing the surface of the samples before grinding, the Ge peaks in the x-ray spectra were no longer present. The
refined lattice constants ($a$ = 4.1384\,\AA, $b$ = 21.838\,\AA, and $c$ = 4.1695\,\AA\, for CeNiGe$_{3}$ and $a$ = 4.0543\,\AA, $b$ =
21.520\,\AA, and $c$ = 4.0617\,\AA\, for YNiGe$_{3}$) are consistent with the earlier studies \cite{Pikul, Bodak}. The structure type, space
group, atomic position, and lattice constants of the single crystals of CeNiGe$_{3}$ and YNiGe$_{3}$ were also confirmed by single crystal x-ray
measurements \cite{Hyunjin}.

Magnetization was measured as a function of temperature from 1.8 to 300\,K and magnetic fields up to 70\,kOe in a Quantum Design (QD) Magnetic
Property Measurement System (MPMS). Heat capacity was measured in a QD Physical Property Measurement System (PPMS) with $^3$He option by the
relaxation method within the temperature range from 0.4 to 50\,K. Electrical resistivity was measured using a standard 4-probe ac ($f$ = 16\,Hz)
technique in a QD PPMS down to 0.4\,K. The plate like samples were cut with a wire-saw (and polished) for resistivity measurements. Electrical
contacts were made to the sample using Epotek H20E silver epoxy to attach Pt wires (50\,$\mu$m diameter). The magnetic field-dependent
resistivity was measured in transverse ((\textbf{I}\,$\parallel$\,\textbf{c})\,$\perp$\,(\textbf{H}\,$\parallel$\,\textbf{a})) configuration for
applied fields up to 90\,kOe.

\section{Results}

\subsection{Magnetic susceptibility}

\begin{figure}
\centering
\includegraphics[width=0.6\linewidth]{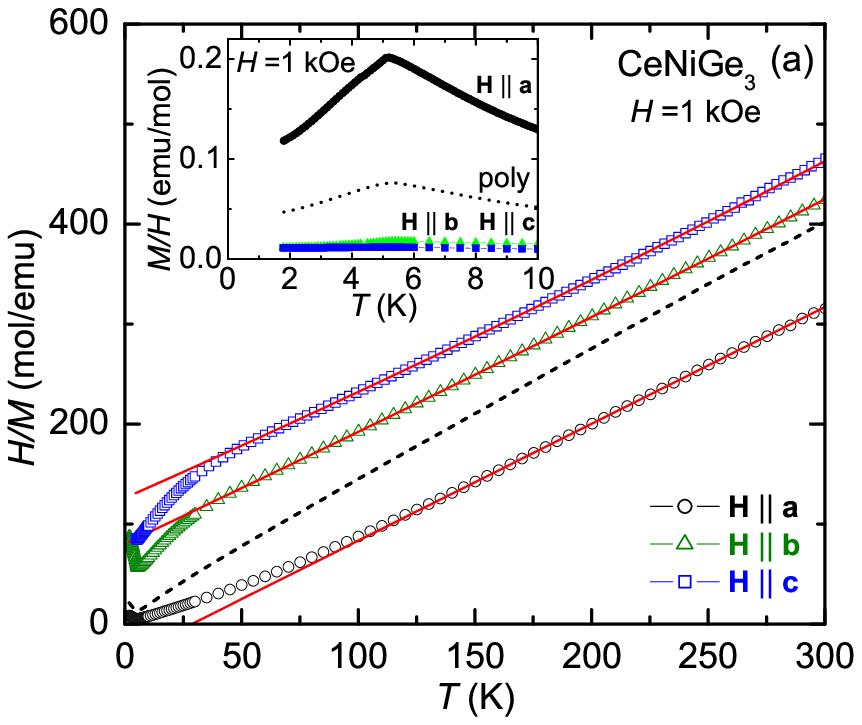}
\includegraphics[width=0.6\linewidth]{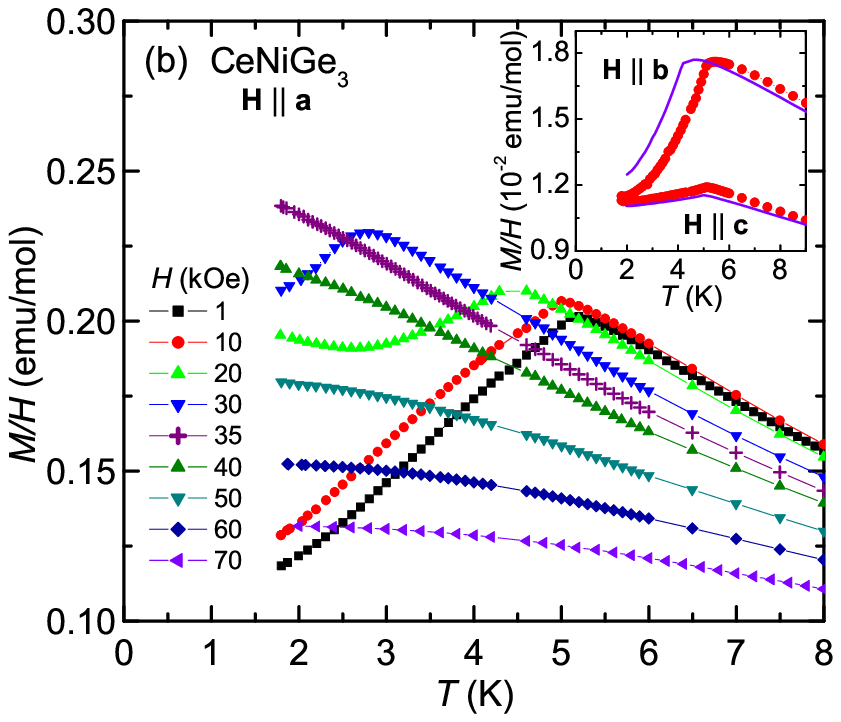}
\caption{(a) Inverse magnetic susceptibility ($H/M(T)$) of CeNiGe$_{3}$ for each principal axis (open symbols) and the polycrystalline average
(dashed line). Solid lines are from the high-temperature Curie-Weiss fits to the data. Inset: Magnetic susceptibility below 10 K for
\textbf{H}\,$\parallel$\,\textbf{a}, polycrystalline average, \textbf{H}\,$\parallel$\,\textbf{b}, and \textbf{H}\,$\parallel$\,\textbf{c} (top
to bottom). (b) $M(T)/H$ of CeNiGe$_{3}$ for \textbf{H}\,$\parallel$\,\textbf{a} at selected magnetic fields. Inset: $M(T)/H$ for
\textbf{H}\,$\parallel$\,\textbf{b} and \textbf{H}\,$\parallel$\,\textbf{c} at $H$ = 10\,kOe (symbols) and 70\,kOe (lines).}
\label{MT}%
\end{figure}

The temperature-dependent magnetic susceptibility ($\chi(T)$ = $M(T)/H$) and inverse magnetic susceptibility, $H$ = 1\,kOe, along each principal
axis are plotted in Fig. \ref{MT} (a). $\chi(T)$ for \textbf{H}\,$\parallel$\,\textbf{a}, which is the magnetic easy axis, shows a sharp peak as a
signature of the AFM order. $\chi(T)$ curves for \textbf{H} $\parallel$ \textbf{b} and \textbf{c} are much smaller than that for \textbf{H}
$\parallel$ \textbf{a}. $\chi(T)$ for each principal axis follows Curie-Weiss law ($\chi(T) = \frac{C}{(T-\theta)}$) above 150\,K.

The strong magnetic anisotropy is also reflected in the anisotropic Weiss temperature ($\theta_{a}$ = 28\,K, $\theta_{b}$ = -60\,K, and
$\theta_{c}$ = -123\,K). Even though the crystal structure of CeNiGe$_{3}$ appears to be ``almost tetragonal" with $a$ and $c$ differing by less
than 1\%, the large difference of $\theta$ between magnetic easy and hard axes results from point symmetry of the Ce site. From the
polycrystalline averaged susceptibility for \textbf{H}\,$\parallel$\,\textbf{a}, \textbf{H}\,$\parallel$\,\textbf{b}, and
\textbf{H}\,$\parallel$\,\textbf{c}, $\chi_{\texttt{poly}}= \frac{1}{3}(\chi_{a}+ \chi_{b} + \chi_{c})$ (dashed line in Fig. \ref{MT} (a)), the
effective moment $\mu_{\texttt{eff}}$ = 2.57\,$\mu_{B}$/Ce$^{3+}$ and the Weiss temperature $\theta_{\texttt{poly}}$ = -21\,K were obtained by
fitting the data to the Curie-Weiss law. The inferred effective moment is close to the full Hund's rule ($J$=5/2) ground-state value,
2.54\,$\mu_{B}$. $\chi(T)$ deviates from the Curie-Weiss law below 150\,K in a manner which suggests that degeneracy of the $J$ = 5/2 of
Ce$^{3+}$ manifold is lifted by CEF (see CEF analysis below).

\begin{figure}
\centering
\includegraphics[width=0.6\linewidth]{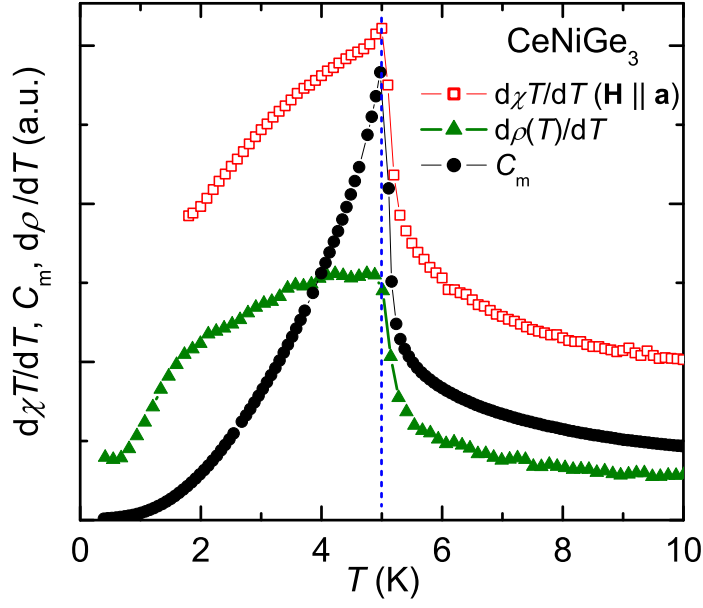}
\caption{Low-temperature d$\chi\cdot$$T$/d$T$ for $H$ = 1\,kOe, $C_{m}$, and d$\rho(T)$/d$T$ for CeNiGe$_{3}$. The antiferromagnetic ordering
temperature marked by dotted-line shows up as a sharp, well-defined peak in all three plots.}
\label{Phase1}%
\end{figure}

The low field, AFM ordering temperature $T_{N}$ = 5.0\,K was determined from the maximum of d$\chi\cdot$$T$/d$T$ as shown in Fig. \ref{Phase1}
\cite{Fisher, Fisher1}, which is consistent with the peak position of specific heat data. The determined AFM ordering temperature is lower than
the previously reported value $T_{N}$ = 5.5\,K \cite{Pikul, Nakashima, Kotegawa, Tateiwa}. The low-temperature $\chi(T)$ data, measured for
\textbf{H}\,$\parallel$\,\textbf{a}, are plotted for various applied fields in Fig. \ref{MT} (b). With increasing magnetic field, $T_{N}$ shifts
to lower temperatures and drops below 2\,K for $H \geq$ 35\,kOe. At higher fields, $\chi(T)$ does not reveal any signature of phase transitions
and instead shows a tendency toward saturation at low temperatures. For \textbf{H}\,$\parallel$\,\textbf{b} and
\textbf{H}\,$\parallel$\,\textbf{c}, however, $T_{N}$ is hardly affected by applied magnetic field as shown in the inset of Fig. \ref{Phase1}; at
$H$ = 70\,kOe, from the maximum of d$\chi\cdot$$T$/d$T$, the determined AFM ordering temperature shifts to $T_{N}$ = 4.1\,K for
\textbf{H}\,$\parallel$\,\textbf{b} and $T_{N}$ = 4.9\,K for \textbf{H}\,$\parallel$\,\textbf{c}.

\subsection{Specific heat}

\begin{figure}
\centering
\includegraphics[width=0.6\linewidth]{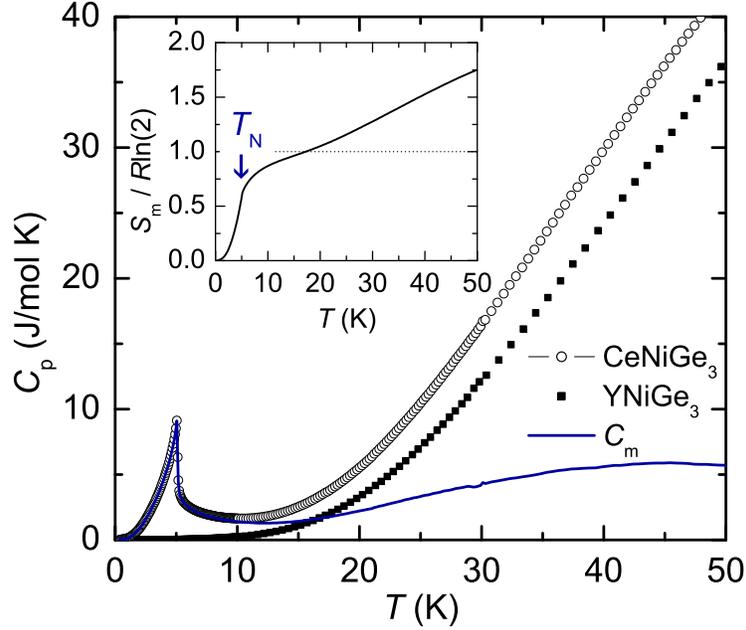}
\caption{Specific heat of CeNiGe$_{3}$ and YNiGe$_{3}$ single crystal and the magnetic specific of CeNiGe$_{3}$ (see text). Inset: Magnetic
entropy divided by $R$ln(2). The vertical arrow indicates the AFM ordering temperature.}
\label{Cp1}%
\end{figure}

\begin{figure}
\centering
\includegraphics[width=0.6\linewidth]{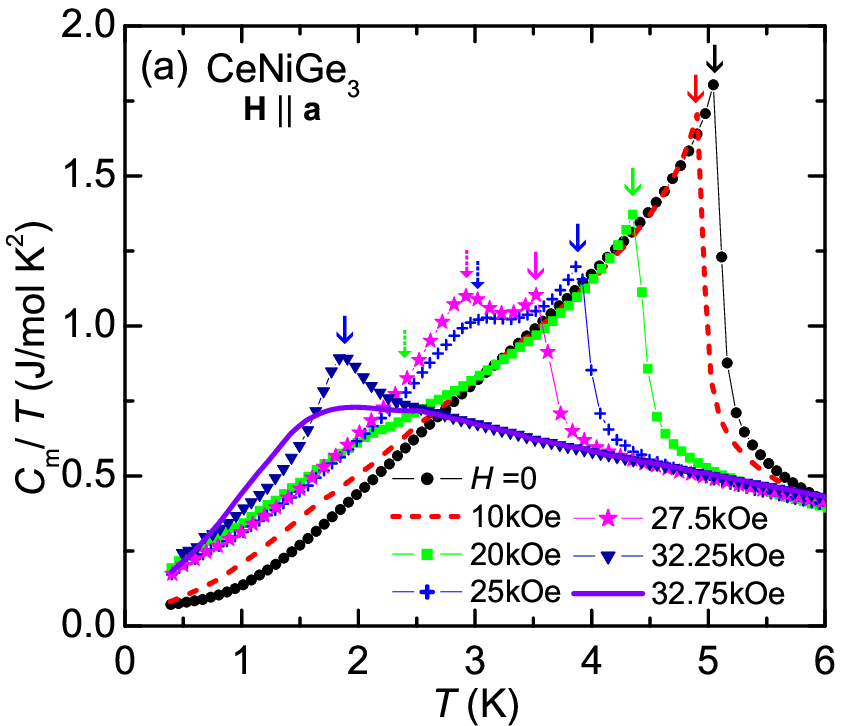}
\includegraphics[width=0.6\linewidth]{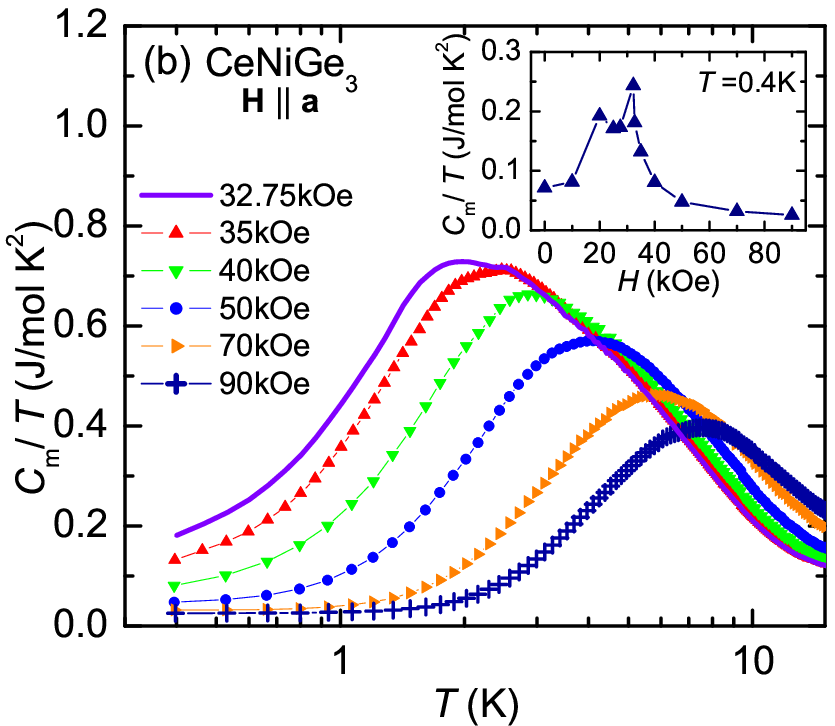}
\caption{Low-temperature magnetic specific heat divided by temperature, $C_{m}/T$, shown for low fields (a) and high fields (b). Larger arrows in
(a) indicate the onset of AFM order and smaller arrows for $H$ = 20, 25, and 27.5\,kOe indicate a low-temperature cusp (see text). The inset of
(b) shows $C_{m}/T$ at 0.4\,K plotted as a function of magnetic field.}
\label{Cp2}%
\end{figure}

Figure \ref{Cp1} shows the temperature-dependent specific heat $C_{p}$ of CeNiGe$_{3}$ and its nonmagnetic, isostructural analogue, YNiGe$_{3}$.
There is a large difference in $C_{p}$ between CeNiGe$_{3}$ and YNiGe$_{3}$ below 50\,K measured. For YNiGe$_{3}$ the electronic specific heat
coefficient ($\gamma$) and Debye temperature ($\Theta_{D}$) were estimated using the relation $C_{p}/T$ = $\gamma$ + $\beta T^{2}$ by plotting
data $C_{p}/T$ vs. $T^{2}$; $\gamma$ = 3.6 mJ/mol$\cdot$K$^{2}$ and $\Theta_{D}$ = 236 K. The specific heat data for CeNiGe$_{3}$ manifests a
clear, sharp, $\lambda$-anomaly with a cusp at 5.0\,K (see Figs. \ref{Cp1} and \ref{Cp2}). Because of the AFM order below 5\,K and the high
temperature, broad feature associated with CEF effect, $\gamma$ and $\Theta_{D}$ of CeNiGe$_{3}$ can not estimated by using the same relation.
$C_{p}/T$ of CeNiGe$_{3}$ attained a value of 76 mJ/mol$\cdot$K$^{2}$ at 0.4\,K, which is larger than that of YNiGe$_{3}$. The data for
YNiGe$_{3}$ are used to estimate the non magnetic contribution to the specific heat of CeNiGe$_{3}$, where we used a mass corrected data
\cite{Bouvier, Blanco} for YNiGe$_{3}$ to subtract from the $C_{p}$ data for CeNiGe$_{3}$. The magnetic specific heat of CeNiGe$_{3}$, $C_{m}$ =
$C_{p}$(CeNiGe$_{3}$) - $C_{p}$(YNiGe$_{3}$), is also plotted in Fig. \ref{Cp1}. The $C_{m}$ data show a broad maximum centered around 45\,K,
indicating that there is a significant magnetic contribution from Ce$^{3+}$ ions above $T_{N}$. This broad peak can be associated with an
electronic Schottky contribution due to the CEF splitting of the Hund's rule ground state multiplet.

The magnetic entropy ($S_{m}$) was inferred by integrating $C_{m}/T$ starting from the lowest temperature measured and plotted in the inset of
Fig. \ref{Cp1}. $S_{m}$ reaches about 60\% of $R$ln(2) at $T_{N}$ and recovers the full doublet, $R$ln(2), entropy by 18\,K. This suggests that
the sharp anomaly at 5.0\,K in specific heat stems from the AFM order in a doublet ground state but the ordered moment is already somewhat
compensated by the Kondo interaction. At $T$ = 50\,K, the recovered $S_{m}$ is less than $R$ln(4), which suggests that the first and second
excited doublets are separated by more than 50\,K from the ground state doublet.

In order to examine the effect of an applied magnetic field on the AFM order, the specific heat was measured with
\textbf{H}\,$\parallel$\,\textbf{a} up to 90\,kOe and is plotted as $C_{m}/T$ vs. $T$ in Fig. \ref{Cp2}. A shift of $T_{N}$ to lower temperatures
is seen for $H <$ 32.5\,kOe and a second, lower temperature, cusp consistent with a second phase transition is observed in the data for 20\,kOe
$\leq$ $H$ $<$ 32.25\,kOe. The position of the low-temperature cusp for $H$ = 20\,kOe can be related to the broad minimum developed in $\chi(T)$
(Fig. \ref{MT} (b)). For $H =$ 32.75\,kOe the magnetic phase transitions are suppressed and instead $C_{m}/T$ exhibits a broad maximum centered
at $\sim$\,2\,K, which marks a cross-over from the AFM to paramagnetic state associated with Brillouin-like saturation of the moments. At higher
fields, this maximum broadens further and moves to higher temperatures as shown in Fig. \ref{Cp2} (b), indicating that the magnetic entropy is
removed at higher temperature for larger applied fields. Such a behavior has been found in Kondo lattice systems under magnetic fields and can be
reasonably reproduced within the Kondo resonance-level model \cite{Schotte} combined with CEF effect. Note that $C_{m}/T$ displays no clear
indication of a low temperature -log($T$), nFL-like behavior \cite{Stewart, Lohneysen} for any of the fields measured. For $H$ = 32.75\,kOe,
$C_{m}/T$ has such a temperature dependence over only a limited temperature range, between 2.5 and 5.5\,K. The electronic specific heat
coefficient $\gamma$, reflecting the effective mass of 4$f$ electrons, as given by $C_{m}/T$ for $T\rightarrow$0, drops quickly as magnetic field
increases. Although at low fields, because of the AFM order, $\gamma$ cannot be formally defined as $C_{m}/T$, $C_{m}/T$ at 0.4\,K is displayed
as a function of field in the inset of Fig. \ref{Cp2} (b), to provide an estimated of $\gamma$. In the paramagnetic state the estimated $\gamma$
values are less than 0.2 J/mole$\cdot$K$^2$. Hence, the strength of the hybridization between 4$f$ and conduction electrons in CeNiGe$_{3}$ is
relatively weak ($f$ electrons in CeNiGe$_{3}$ are localized) and it can be destroyed easily by the application of magnetic field, which is
evidenced by the value $\gamma$ = 25 mJ/mole$\cdot$K$^2$ for $H$ = 90\,kOe.

\subsection{Resistivity}

\begin{figure}
\centering
\includegraphics[width=0.6\linewidth]{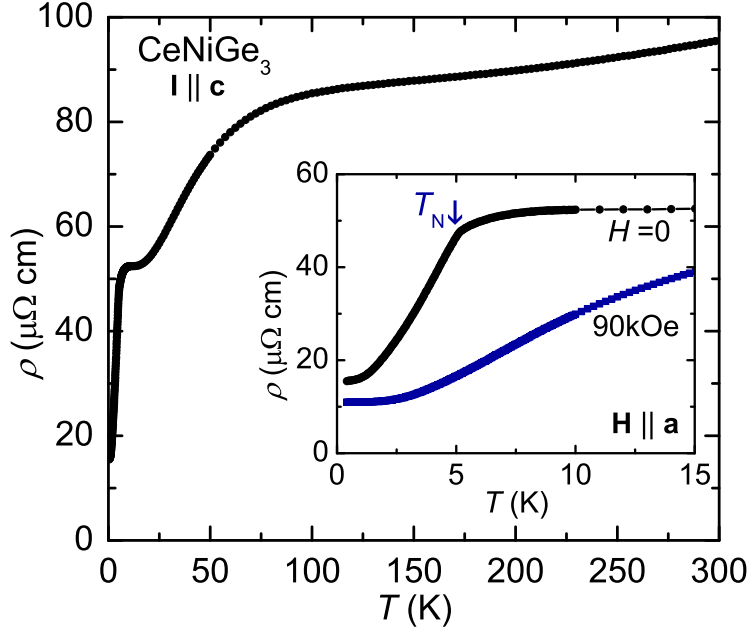}
\caption{Temperature-dependent electrical resistivity ($\rho(T)$) of CeNiGe$_{3}$. Inset: $\rho(T)$ measured at $H$ = 0 and 90\,kOe for magnetic
field applied along \textbf{H}\,$\parallel$\,\textbf{a}. The onset of AFM order for $H$ = 0 is indicated by the arrow.}
\label{RT}%
\end{figure}

The temperature-dependent resistivity data ($\rho(T)$) of CeNiGe$_{3}$ are shown in Fig. \ref{RT} down to 0.4\,K in zero field. There are two
broad features in $\rho(T)$ around at 10\,K and 100\,K, the lower one being primarily related to Kondo scattering and the higher one primarily
being associated with thermal population of CEF levels. The inset of Fig. \ref{RT} shows the zero field $\rho(T)$ below 15\,K, together with
90\,kOe data for magnetic field \textbf{H}\,$\parallel$\,\textbf{a}. The AFM transition manifests itself as a precipitous drop in $\rho(T)$ for
$T < T_{N}$.

From the results of our thermodynamic and transport measurements, the AFM ordering temperature can be clearly determined in $\chi(T)$, $C_{p}(T)$,
and $\rho(T)$. Figure \ref{Phase1} shows d$\chi\cdot T$/d$T$, $C_{m}(T)$, and d$\rho(T)$/d$T$ curves \cite{Fisher, Fisher1}. There is an excellent
agreement among d$\chi\cdot T$/d$T$ = 5.0(0)\,K, $C_{m}(T)$ = 4.9(7)\,K, and d$\rho(T)$/d$T$ = 4.9(4)\,K. The same analysis to determine phase
transition temperatures has been applied to more complex and/or local moment system such as $R$Ni$_{2}$B$_{2}$C ($R$ = Dy and Ho) \cite{Ribeiro}
and shows that the reliable phase transition temperatures can be obtained from thermodynamic and transport measurements.

%Below $T_{N}$, $\rho(T)$ indicates a $T^{1.5}$-dependence down to 1.5 K and further cooling shows an exponential-dependence on temperature, where
%no evidence for Fermi liquid (FL) behavior ($\Delta\rho = AT^{n}$ with $n$ = 2) is observed. Thus, magnetic contribution might be significant
%below $T_{N}$. It was reported from earlier study on polycrystal samples that below $T_{N}$ $\rho(T)$ and $C_{m}$ is well described by
%considering the spin wave contribution \cite{Pikul}.

\begin{figure}
\centering
\includegraphics[width=0.6\linewidth]{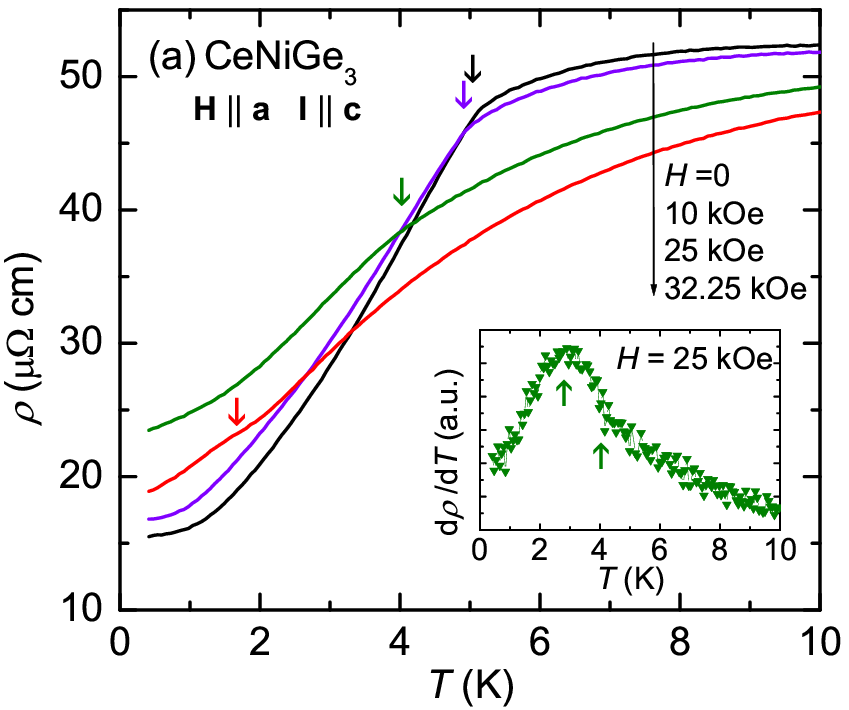}
\includegraphics[width=0.6\linewidth]{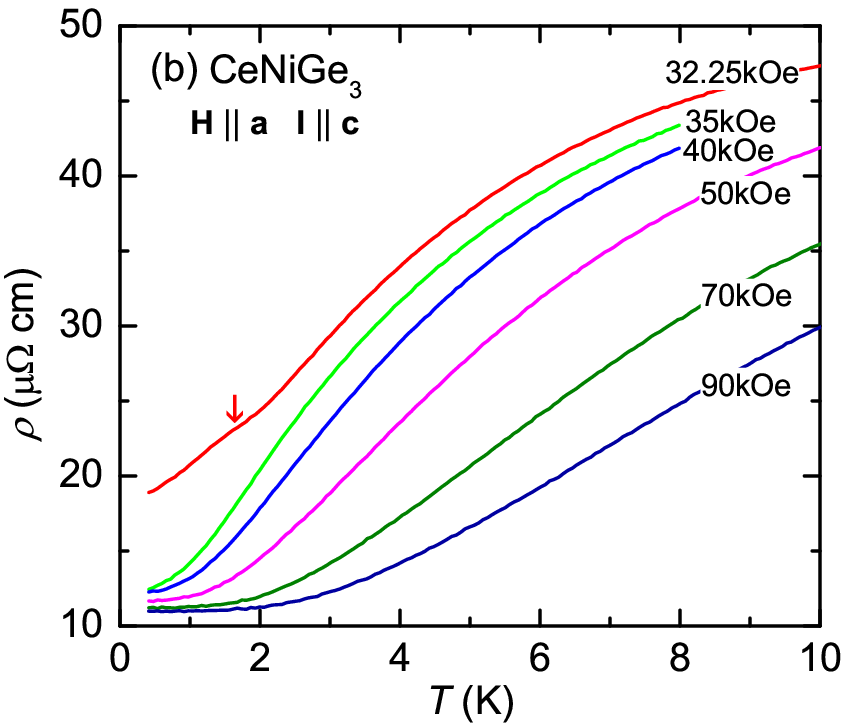}
\caption{Low-temperature electrical resistivity, $\rho(T)$, at selected magnetic fields applied along \textbf{H}\,$\parallel$\,\textbf{a}. (a)
$H$ = 0, 10, 25, and 32.25\,kOe. Inset: d$\rho(T)$/d$T$ for $H$ = 25\,kOe. Arrows indicate a maximum and slope change in d$\rho(T)$/d$T$. (b) $H$
= 32.25, 35, 40, 50, 70, and 90\,kOe. Arrows indicate the onset of AFM ordering (see text).}
\label{RT1}%
\end{figure}

As magnetic field increases, a gradual suppression of AFM order is observed in $\rho(T)$. The determined AFM phase transition temperatures from
d$\rho(T)$/d$T$ are indicated by arrows in Fig. \ref{RT1} (a). The low-temperature $\rho(T)$ for intermediate fields, for instance $H$ = 32.25 kOe
data, reveals more scattering at the phase transition temperature than that in zero field which shows a sharp drop due to loss of spin disorder
scattering. The inset of Fig. \ref{RT1} (a) shows the d$\rho(T)$/d$T$ curve for $H$ = 25\,kOe in which two anomalies can be seen around $\sim$
2.8\,K as a maximum and $\sim$ 3.9\,K as a slope change. These temperatures are consistent with the cusp and sharp peak position developed in
$C_{m}(T)$ (Fig. \ref{Cp2}). At higher fields, FL behavior ($\rho\propto T^{2}$) is found over a limited range of temperature and field. At very
low temperatures and high fields, $\rho(T)$ becomes flat, revealing an anomalous $T^{n}$-dependence with $n>2$.

\subsection{Magnetization and Magnetoresistance}

\begin{figure}
\centering
\includegraphics[width=0.6\linewidth]{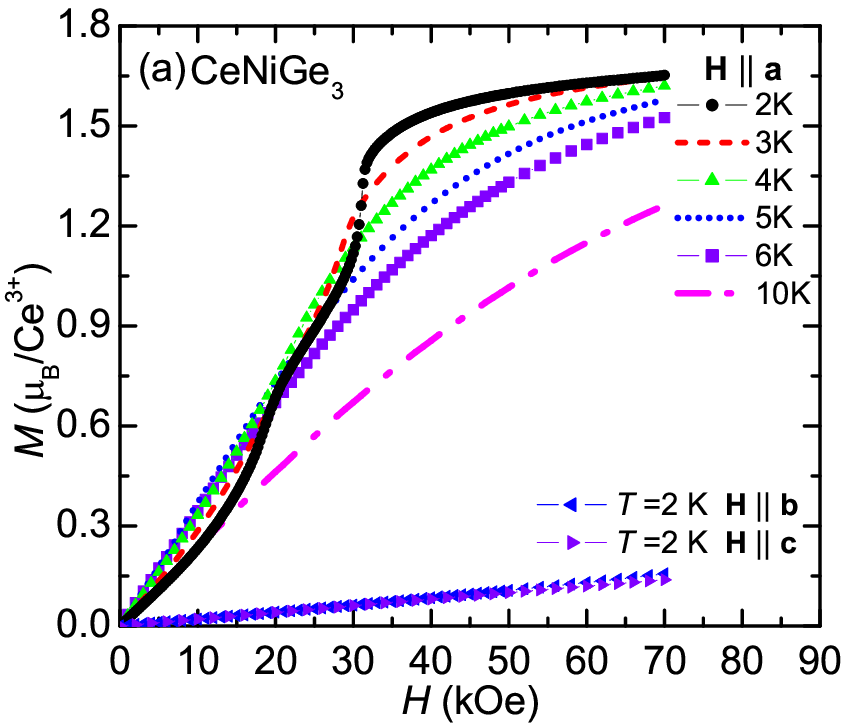}
\includegraphics[width=0.6\linewidth]{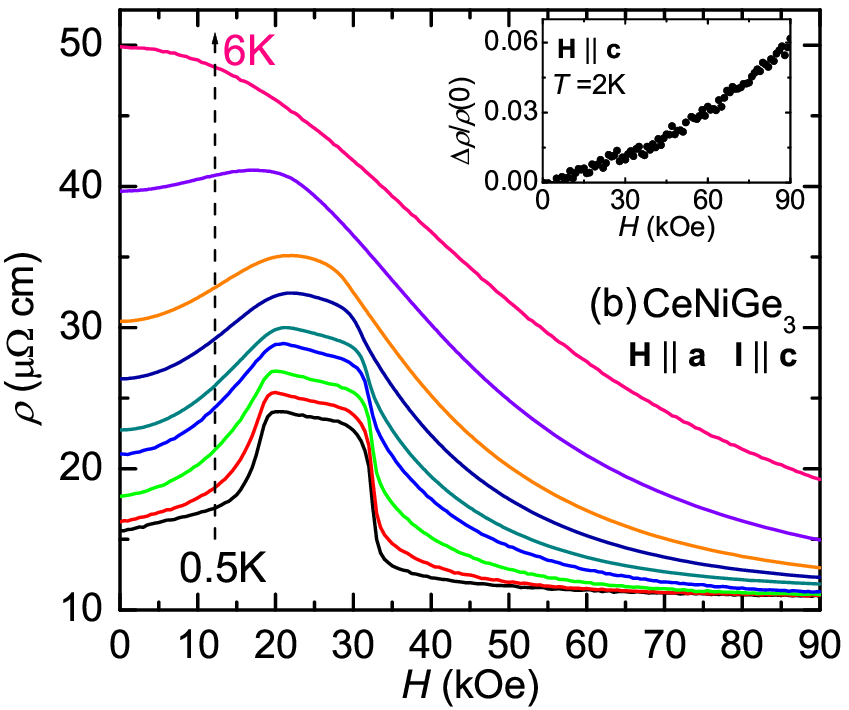}
\caption{(a) Magnetization isotherms for \textbf{H}\,$\parallel$\,\textbf{a} at $T$ = 2, 3, 4, 5, 6, and 10\,K and for
\textbf{H}\,$\parallel$\,\textbf{b} and \textbf{H}\,$\parallel$\,\textbf{c} at 2\,K. (b) Electrical resistivity as a function of magnetic field
for \textbf{H}\,$\parallel$\,\textbf{a} at $T$ = 0.5, 1, 1.5, 2, 2.25, 2.75, 3.25, 4.25, and 6\,K. Inset: $\rho(H)$ for
\textbf{H}\,$\parallel$\,\textbf{c} at 2\,K, plotted as $\Delta\rho$/$\rho$(0) vs. $H$.}
\label{MHRH}%
\end{figure}

\begin{figure}
\centering
\includegraphics[width=0.6\linewidth]{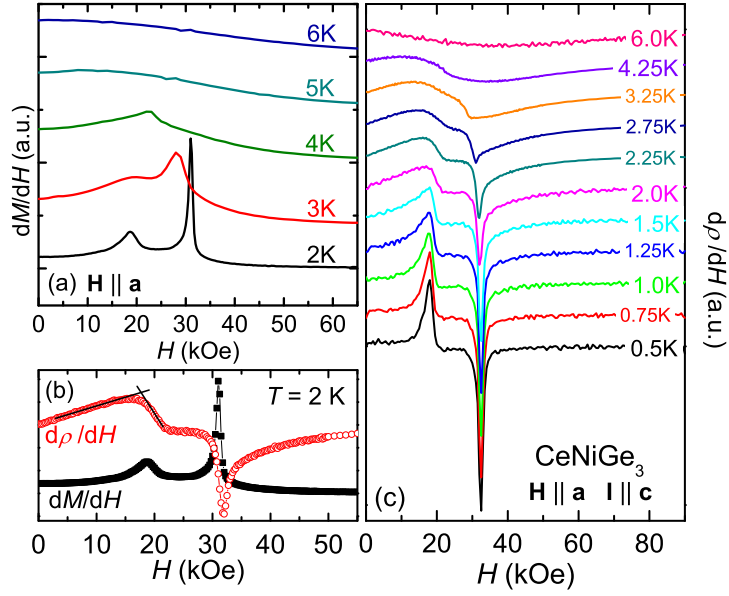}
\caption{(a) d$M(H)$/d$H$ for \textbf{H}\,$\parallel$\,\textbf{a} at $T$ = 2, 3, 4, 5, and 6\,K. (b) d$\rho(H)$/d$H$ and d$M(H)$/d$H$ at $T$ =
2\,K. Solid lines are guide to eye. (c) d$\rho(H)$/d$H$ (\textbf{H}\,$\parallel$\,\textbf{a}) at selected temperatures between 0.5 and 6\,K.}
\label{Phase2}%
\end{figure}

The magnetic field-dependent magnetization ($M(H)$) and resistivity ($\rho(H)$) shed further light on the low-temperature magnetic states of
CeNiGe$_{3}$, as shown in Fig. \ref{MHRH}. The observed magnetization isotherms at 2\,K are highly anisotropic between
\textbf{H}\,$\parallel$\,\textbf{a} and $H$ in the \textbf{bc}-plane as seen in Fig. \ref{MHRH} (a). At 2\,K, $M(H)$ for
\textbf{H}\,$\parallel$\,\textbf{b} and \textbf{H}\,$\parallel$\,\textbf{c} increases almost linearly up to about 0.15 $\mu_{B}$/Ce$^{3+}$ at
70\,kOe. On the other hand, $M(H)$ along \textbf{H}\,$\parallel$\,\textbf{a} linearly increases below 10\,kOe and then, undergoes two MMTs and
appears to be approaching saturation at 70\,kOe with a moment of about 1.65 $\mu_{B}$/Ce$^{3+}$. This value is much higher, by factor of
$\sim$\,4, than in the previous polycrystalline sample study \cite{Pikul} but lower than the theoretical value of 2.14\,$\mu_{B}$ for the
saturated moment of free Ce$^{3+}$ ions. This increase in $M(H)$ takes place via two MMTs at 18.5 and 31\,kOe, which are seen as steps in $M(H)$
and also clearly indicated in the d$M(H)$/d$H$ analysis. Figure \ref{Phase2} (a) shows d$M(H)$/d$H$ at selected temperatures. Two peaks in
d$M(H)$/d$H$ curves, corresponding to MMTs, are no longer seen for $T$ $\geq$ 5\,K. For \textbf{H}\,$\parallel$\,\textbf{b} and
\textbf{H}\,$\parallel$\,\textbf{c}, no MMTs are observed below 70\,kOe and down to 2\,K.

\begin{figure}
\centering
\includegraphics[width=0.6\linewidth]{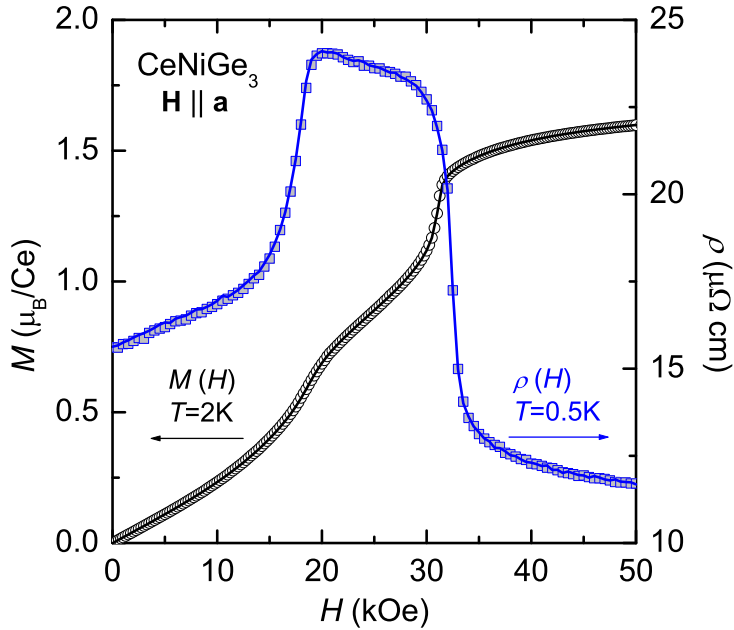}
\caption{Magnetization isotherms (left axis) at $T$ = 2\,K and magnetoresistivities (right axis) at $T$ = 0.5\,K for
\textbf{H}\,$\parallel$\,\textbf{a}. Symbols and lines are taken the data with increasing and decreasing magnetic fields, respectively, at fixed
temperature.}
\label{MHRH1}%
\end{figure}

$\rho(H)$ measurements provide orthogonal cuts through the $H-T$ phase diagram and shed light on some of the features observed in $\rho(T)$.
Figure \ref{MHRH} (b) presents $\rho(H)$ for \textbf{H}\,$\parallel$\,\textbf{a} below 6\,K. Dramatic changes in $\rho(H)$ for $T$ $\leq$ 4.25\,K
reveal the two MMTs, while a monotonically decreasing $\rho(H)$ is seen in the 6\,K curve up to 90\,kOe. Two MMTs are clearly revealed in the
d$\rho(H)$/d$H$ curve as sharp peaks. For comparison d$M(H)$/d$H$ and d$\rho(H)$/d$H$ curve at 2\,K is plotted in Figure \ref{Phase2} (b). Two
MMTs fields revealed at 18 and 31.5\,kOe in d$\rho(H)$/d$H$ are in agreement with the features developed at 18.5\,kOe and 31\,kOe in d$M(H)$/d$H$.
Below 2\,K two MMTs are almost temperature independent, while above 2\,K the second MMT shifts to lower field as temperature increases. Note that
the pronounced steps in $\rho(H)$ were not observed in the polycrystalline sample study \cite{Pikul}. Magnetization at $T$ = 2\,K and
magnetoresistivity vs. field loops from 0 to 50\,kOe at $T$ = 0.5\,K are shown in Fig. \ref{MHRH1}. Although CeNiGe$_{3}$ show sharp MMTs, $M(H)$
indicates no hysteresis down to 2\,K and no hysteresis is seen in $\rho(H)$ measurements at 0.5\,K. In the inset of Fig. \ref{MHRH} (b) $\rho(H)$
for \textbf{H}\,$\parallel$\,\textbf{c} at 2\,K is plotted as $\Delta\rho/\rho$(0) vs. $H$. As magnetic field increases $\rho(H)$ increases
monotonically without step-like change shown for \textbf{H}\,$\parallel$\,\textbf{a}, which is consistent with $M(H)$ results for
\textbf{H}\,$\parallel$\,\textbf{c}. Below $T_{N}$, the observed magnetoresistance (MR) in CeNiGe$_{3}$ is very similar to that in a tetragonal
CeAuSb$_{2}$ \cite{Balicas} and a isostructural YbNiSi$_{3}$ \cite{Budko}.

%In the AFM state, as magnetic field increases parallel to one of the sublattice, the fluctuations of the spins can be suppressed while the other
%sublattice may increase. Thus, the magnetoresistance (MR) of AFM material can be changed from positive to negative passing through zero, because
%the fluctuations of both sublattice spins are reflected in MR \cite{Yamada}. The applied magnetic field eliminates the AFM spin interactions,
%aligning gradually Ce$^{3+}$ spins and inducing the saturated paramagnetic state (SPM, field induced ferromagnetic state).

\section{Discussion}

\begin{figure}
\centering
\includegraphics[width=0.6\linewidth]{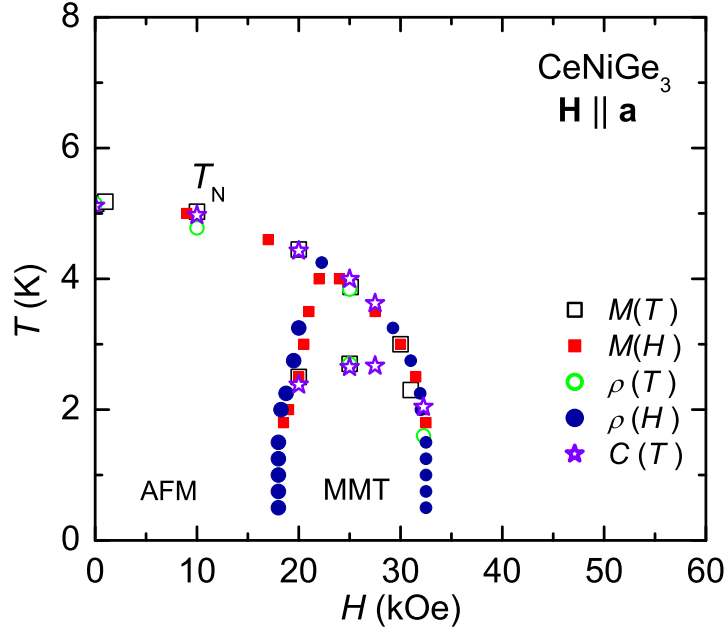}
\caption{$H-T$ phase diagram of CeNiGe$_{3}$ for \textbf{H}\,$\parallel$\,\textbf{a}. Symbols are taken from temperature and field sweeps using
d$\chi\cdot T$/d$T$, d$\rho(T)$/d$T$, $C_{m}$, d$M(H)$/d$H$, and d$\rho(H)$/d$H$ analysis.}
\label{Phase}%
\end{figure}

In Fig. \ref{Phase} the phase transition temperatures and fields determined from magnetization, resistivity, and specific heat measurements are
plotted in the $H-T$ plane, where symbols are extracted from d$\chi\cdot T$/d$T$, d$\rho(T)$/d$T$, d$M(H)$/d$H$, d$\rho(H)$/d$H$ analysis, and
maximum of $C_{m}(T)$. The phase transition can be traced by a line which is connected from $T_{N}$ = 5.0\,K at $H$ = 0 to the critical field
$H_{c}$ = 32.5\,kOe at $T$ = 0.5\,K. An additional phase boundary emerges below 4\,K around 22\,kOe. This phase boundary depends weakly on field.
Between these two phase boundaries (20\,kOe $< H <$ 30\,kOe), a nearly temperature-independent phase line seems to exist, which corresponds to
the low-temperature cusp observed in $C_{m}$ (Fig. \ref{Cp2}) and the maximum in d$\rho(T)$/d$T$. At present, it is not clear the low-temperature
cusp in $C_{m}$ is a indication of phase transition or is associated with a excitation of spin wave. As shown in the phase diagram the critical
temperature determined from temperature sweeps track well the critical field determined from field sweeps.

$T_{N}$ gradually shifts to lower temperature in applied fields up to $\sim$ 30\,kOe. It then decreases quickly and seems to be zero near $H_{c}$
$\sim$ 32.5\,kOe. In spite of the abrupt changes in the slope of $\rho(H)$, no hysteresis was observed at $H_{c}$. Above $H_{c}$ the saturated
paramagnetic state with near fully saturated moment is induced. Since a pronounced drop of the resistivity at $H_{c}$ sharpens as temperature
decrease, the transition associated with the MMT may well be first order below 0.4\,K. In Ref. \cite{Pikul} the phase diagram based on
polycrystalline samples indicates three distinct magnetic phase boundaries. The two phase boundaries between regions I, II, and III in Ref.
\cite{Pikul} are consistent with the phase lines for \textbf{H} $\parallel$ \textbf{a} in this study. However, the third phase boundary between
region III and paramagnetic phase in Ref. \cite{Pikul} is not present in Fig. \ref{Phase}. From a simple point of view, since CeNiGe$_{3}$ is
highly magnetically anisotropic, the third phase boundary seen in the polycrystalline samples may primarily reflect that for the magnetic hard
axes \textbf{b} or \textbf{c}. We have been measured $\chi(T)$ for magnetic field applied along magnetic hard axes (see Fig. \ref{MT}). For $H$ =
70\,kOe, $\chi(T)$ reveals a shift of the AFM ordering temperature from 5.0 to 4.2\,K for \textbf{H}\,$\parallel$\,\textbf{b} and to 4.9\,K for
\textbf{H}\,$\parallel$\,\textbf{c}, which is close to the third phase boundary in Ref. \cite{Pikul}. The constructed $H-T$ phase diagram for
\textbf{H}\,$\parallel$\,\textbf{a} is qualitatively in agreement with the phase diagram from previous polycrystalline sample study \cite{Pikul}.
Hence, although polycrystalline samples can show intrinsic property of materials with directional averaged value, single crystals are needed for
reliable $H-T$ phase diagrams of magnetically anisotropic compounds.

CeNiGe$_{3}$ manifests a large magnetic anisotropy and the specific heat shows a broad peak centered around 45\,K in zero field, both of which
are presumably associated with the CEF splitting of the Ce Hund's rule ground state multiplet. In order to better understand the salient energy
scales for this system, these data were analyzed on the basis of the CEF (point charge) model. The CEF Hamiltonian for the Ce$^{3+}$ ion
($J$=5/2) in orthorhombic point symmetry, can be written as $H_{\texttt{CEF}}$ = $B^{0}_{2} O^{0}_{2}$ + $B^{2}_{2} O^{2}_{2}$ + $B^{0}_{4}
O^{0}_{4}$ + $B^{2}_{4} O^{2}_{4}$ + $B^{4}_{4} O^{4}_{4}$. Where $O^{n}_{m}$ and $B^{n}_{m}$ are Steven's operators and crystal field
parameters, respectively \cite{Stevens, Hutchings}. In the paramagnetic state, the magnetic properties of the material are determined by CEF plus
Zeeman interaction ($H_{\texttt{Z}}$ = $g\mu_{B}J H$), thus the total Hamiltonian $H = H_{\texttt{CEF}} + H_{\texttt{Z}}$ was used for the
analysis of the present data.

\begin{figure*}
\centering
\includegraphics[width=0.33\linewidth]{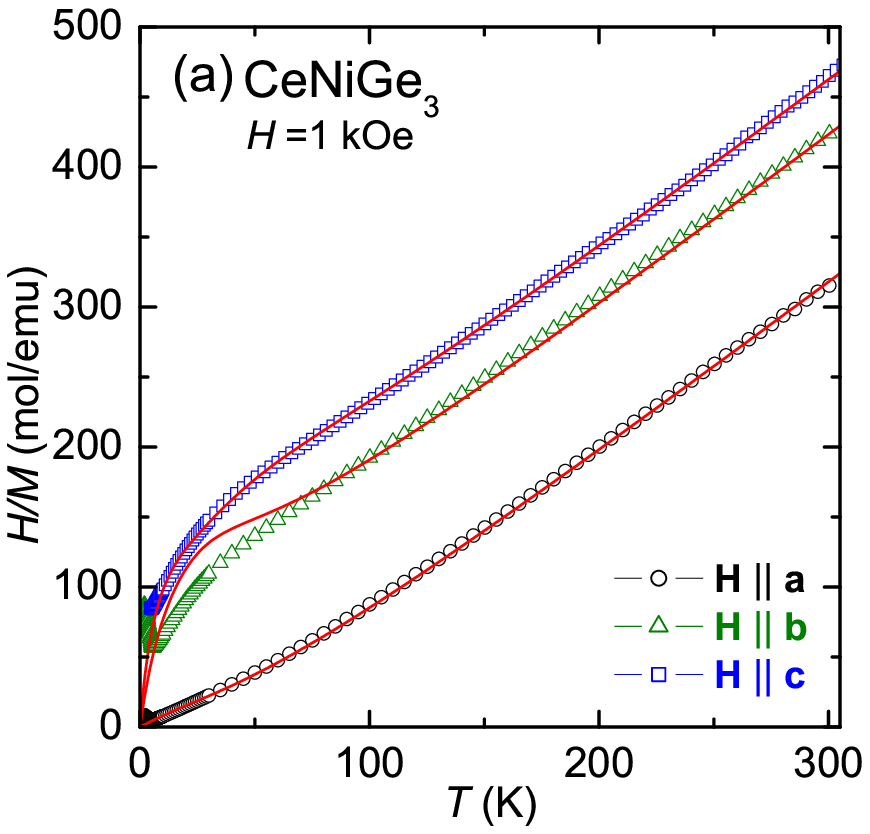}\includegraphics[width=0.33\linewidth]{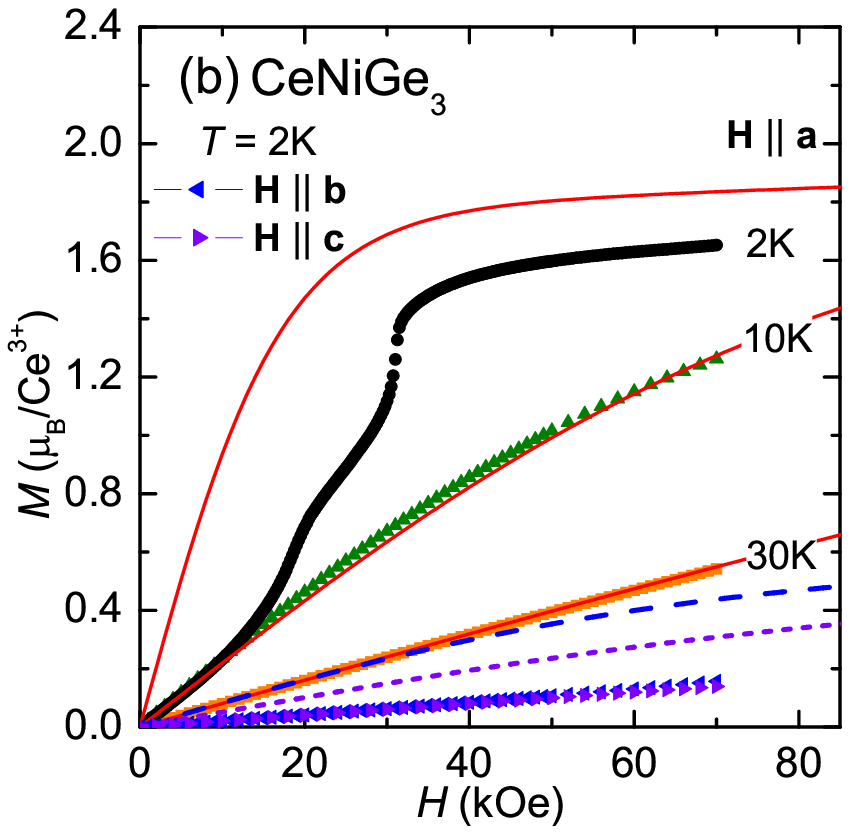}\includegraphics[width=0.33\linewidth]{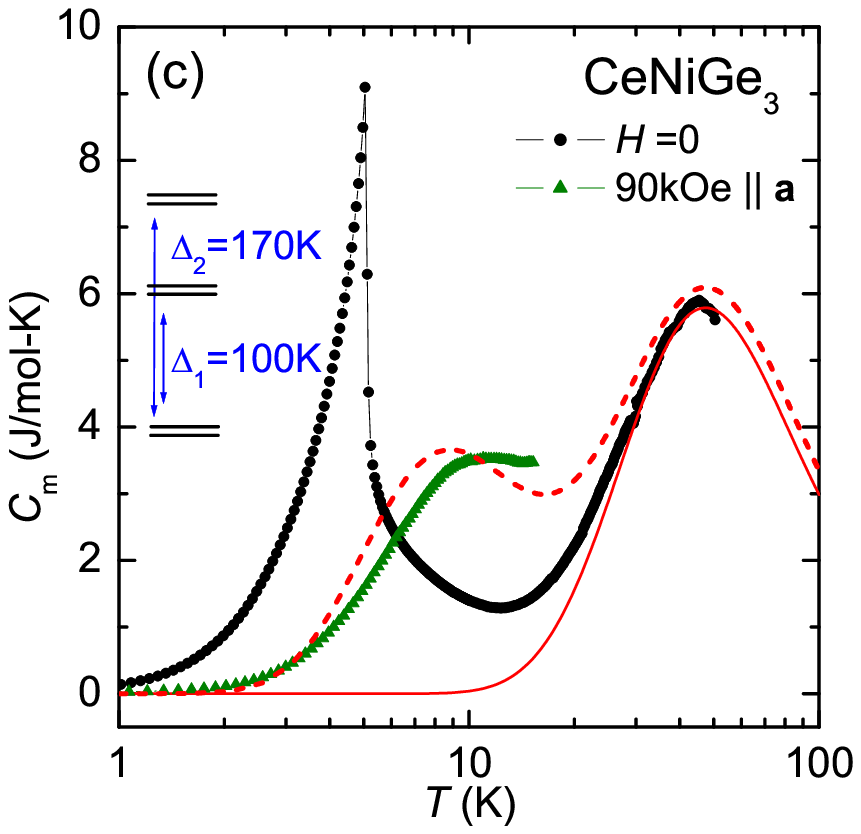}
\caption{(a) Inverse magnetic susceptibility (open symbols) and calculated curves based on the CEF model (solid lines). (b) Magnetization
isotherms for \textbf{H}\,$\parallel$\,\textbf{a} (2, 10, 30\,K), \textbf{H}\,$\parallel$\,\textbf{b} (2\,K) and
\textbf{H}\,$\parallel$\,\textbf{c} (2\,K). Solid lines are the calculated curves based on the CEF model for \textbf{H}\,$\parallel$\,\textbf{a}
and $T$ = 2, 10, 30\,K. Dashed (dotted) line is the calculated curve for \textbf{H}\,$\parallel$\,\textbf{b}
(\textbf{H}\,$\parallel$\,\textbf{c}). (c) Magnetic specific heat at $H$ =0 and 90\,kOe. Solid and dashed lines are the calculated magnetic
specific heat curves based on the CEF model at $H$ = 0 and 90\,kOe, respectively, for \textbf{H}\,$\parallel$\,\textbf{a}. The given energy level
scheme, shown on the left side, represents that the first and second excited state energy level of the Kramers doublets are separated by
$\Delta_{1}$ $\sim$ 100\,K and $\Delta_{2}$ $\sim$ 170\,K, respectively, from the ground state doublet.}
\label{CEF}%
\end{figure*}

The CEF parameters were determined by reproducing the measured specific heat, magnetic susceptiblity, and magnetization isotherms with calculated
curves shown as solid line in Fig. \ref{CEF}. The obtained CEF parameters and the CEF level scheme are summarized in Table \ref{table1} and the
zero field energy level scheme is shown in Fig. \ref{CEF} (c). In the paramagnetic state, $\chi(T)$ and $M(H)$ are satisfactorily reproduced by
the calculated values as shown in Figs. \ref{CEF} (a) and (b). Although $M(H)$ data at 2\,K are not exactly reproduced, the anisotropy between
\textbf{H}\,$\parallel$\,\textbf{a} and \textbf{H}\,$\parallel$\,\textbf{b} and \textbf{c} is reasonable. The two MMTs, of course, are not
captured by single ion, CEF effect. Thus, in addition to the CEF effect, other interactions should be taken into consideration to reproduce the
MMT behavior in CeNiGe$_{3}$. In similar manner, the magnetic ordering seen as a $\lambda$-like anomaly in the zero field $C_{p}$ is not captured
in the CEF model, but the high field splitting of the ground state doublet is. At high temperatures $C_{m}$ is well reproduced by the CEF model as
shown in Fig. \ref{CEF} (c). The energy level splitting $\Delta_{1}$ = 100\,K from the ground state doublet to the first excited doublet is
consistent with the entropy change shown in Fig. \ref{Cp1}. Furthermore, for $H$ = 90\,kOe, the low temperature broad peak is reproduced, giving
rise to a split of ground state double via Zeeman interaction.

\begin{figure}
\centering
\includegraphics[width=0.6\linewidth]{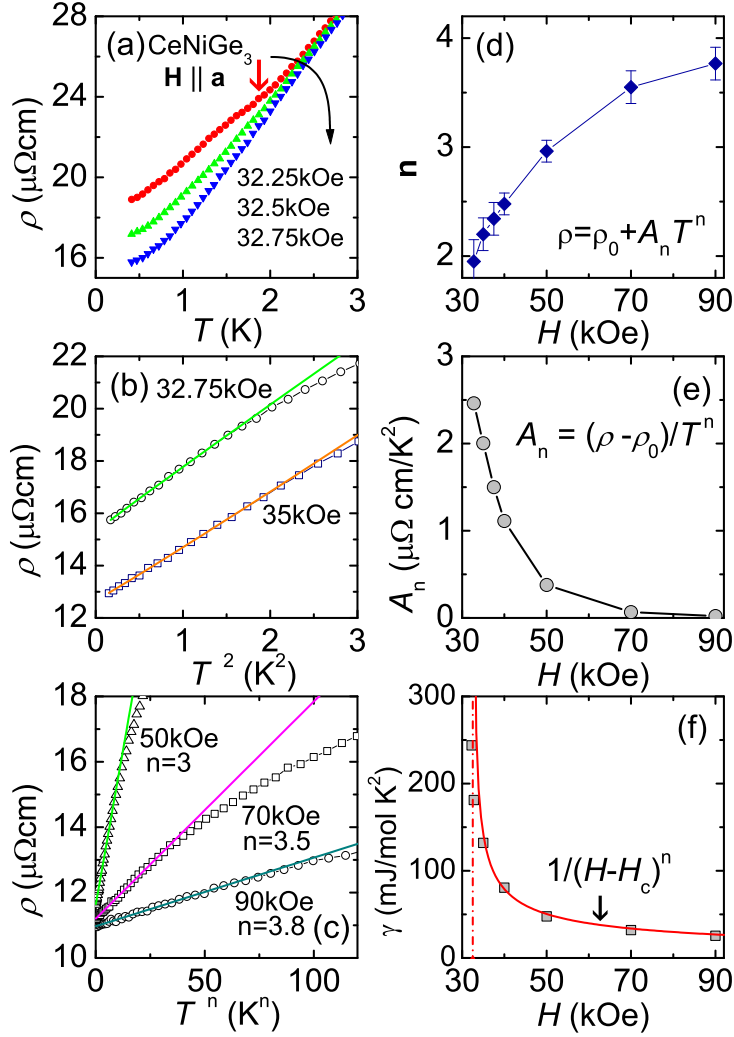}
\caption{(a) Low-temperature electrical resistivity at 32.25, 32.5, and 32.72\,kOe for \textbf{H}$\parallel$\textbf{a}. Vertical arrow indicates
the onset of the AFM phase transition. (b) Resistivity as a function of $T^{2}$ for $H$ = 32.75 and 35\,kOe. (c) Resistivity as a function of
$T^{n}$, $\rho(T)=\rho_{0}+A_{n}T^{n}$, at $H$ = 50\,kOe ($n$ = 3), 70\,kOe ($n$ = 3.5) and 90\,kOe ($n$ = 3.8). (d) Exponent $n$ (e) coefficient
$A_{n}$ as a function of magnetic field. (f) $C_{m}/T$ at $T$ = 0.4\,K (inset, Fig. \ref{Cp2}). The solid line represents a fit of the equation,
$C_{m}/T$ $\propto$ 1/$(H-H_{c})^{\beta}$, performed between 35 and 90\,kOe with $n$ = 0.52 and $H_{c}$ = 32.5\,kOe. The vertical dash-dotted
line indicates $H_{c}$.}
\label{Power}%
\end{figure}

Turning now to the low temperature resistivity, it is expected that FL behavior should be recovered for $H > H_{c}$. Figure \ref{Power} shows
that, for selected magnetic fields, $\rho(T)$ for CeNiGe$_{3}$ varies as $\rho(T)$ = $\rho_{0}$ + $AT^{n}$, with $n \geq 2$ below 2\,K. The
vertical arrow in Fig. \ref{Power} (a) indicates the onset of AFM order, below which a small enhancement of $\rho(T)$ is observed rather than a
drop due to the loss of spin disorder scattering. As magnetic field increase from 32.25 to 32.75\,kOe the AFM order is fully suppressed. In the
vicinity of the AFM to PM phase boundary a quadratic temperature dependence in $\rho(T)$ emerges (see Fig. \ref{Power} (b)). A FL behavior,
$n\,=\,2$, is obtained just above $H_{c}$, but only over a limited range of temperature and magnetic field. At low temperatures, $H > H_{c}$,
$\rho(T)$ displays an anomalous $T^{n}$ dependence with $n >$ 2, indicative of deviation from FL behavior, as shown in Fig. \ref{Power} (c). This
suggests that an additional scattering mechanism in $\rho(T)$ needs to be considered in the saturated paramagnetic (spin polarized) states. The
exponent, $n$, gradually increases as magnetic field increases, reaching $n$ = 3.8 at $H$ = 90\,kOe (Fig. \ref{Power} (d)), where the coefficient,
$A_{n}$ = ($\rho(T)-\rho_{0}$)/$T^{n}$, corresponding to each $n$ is plotted in Fig. \ref{Power} (e). The $A_{n}$ value quickly drops with
increasing magnetic fields for $H\,>\,H_{c}$.

In general, when $T_{N}$, tuned by external magnetic field, is suppressed to $T \rightarrow 0$, pronounced nFL properties are observed in the
vicinity of QCP in HF AFM systems such as YbRh$_{2}$Si$_{2}$ \cite{Gegenwart, Custers}, CeAuSb$_{2}$ \cite{Balicas}, YbAgGe \cite{Budko1,
Niklowitz, Tokiwa}. For CeNiGe$_{3}$ there is no clear indication of nFL behavior, $\Delta\rho = AT^{n}$ with $n <$ 2 and
$C/T\propto$\,-log($T$), in the vicinity of $H_{c}$ down to 0.4\,K. Therefore, although the AFM order in CeNiGe$_{3}$ can be suppressed by an
applied magnetic field, a classical QCP is not obtained as $T_{N} \rightarrow 0$. This may arise due to a weak hybridization between 4$f$ and
conduction electrons and the possible first order nature of this phase transition for $T <$ 0.4\,K and $H$ $\sim$ $H_{c}$, i.e. Ce in this system
is much more local moment like and the $H-T$ phase diagram is closer to non-hybridizing rare-earth such as $R$AgSb$_{2}$ or $R$AgGe system ($R$ =
rare-earth) \cite{Myers, Morosan}. For many field tuned QCP systems \cite{Gegenwart, Balicas, Lohneysen}, when approaching the QCP from the
paramagnetic state, the coefficient $A$ of the $T^{2}$-dependence of resistivity has a 1/$(H-H_{c})^{\beta}$ scaling form, where $H_{c}$ is the
critical field and $\beta$ is the exponent. Since $\rho(T)$ of CeNiGe$_{3}$ shows a $T^{2}$-dependence only over a limited temperature and field
range, this scaling property can not tested in this system. Interestingly, however, a 1/$(H-H_{c})^{0.52}$ divergence for $H > H_{c}$ is observed
from the field dependence of $C_{m}/T$ at 0.4\,K. The solid line in Fig. \ref{Power} (f) represents a fit of the scaling form performed for $H
\geq$ 35\,kOe with $H_{c}$ = 32.5 $\pm$ 0.5\,kOe and $\beta$ = 0.52 $\pm$ 0.1, implying a divergence of the quasi-particle mass approaching
$H_{c}$. By assuming $A \propto \gamma^{2}$ \cite{Kadowaki} holds in the paramagentic state, this implies that the field dependence $A$ can show a
singular scattering near $H_{c}$ following $\sim$ 1/$(H-H_{c})$. In order to address this issue, it will be necessary to measure the resisitivity
below 0.4\,K. It is interesting to note that the $n\,>$ 2 deviation from FL behavior is also seen in CeAuSb$_{2}$ \cite{Balicas} and YbNiSi$_{3}$
\cite{Budko}. Both compounds show very similar $H-T$ phase diagrams with two MMTs, emerging when $T_{N}$ is lowered by applied magnetic field.
The observed MR response through the MMTs is also quite similar. The anomalous metallic behavior in the saturated paramagnetic state induced by
magnetic field is reflected by $T^{3}$ dependence for CeAuSb$_{2}$ and $T^{n}$ with $n >$ 2 for YbNiSi$_{3}$ at low temperatures.

\section{Summary}

We have studied the thermal, magnetic, and electrical properties of the single crystals of CeNiGe$_{3}$ by means of the specific heat,
magnetization, and electrical resistivity. CeNiGe$_{3}$ manifest a large magnetic anisotropy with the crystallographic \textbf{a}-axis being the
magnetic easy axis. In the paramagnetic phase, the observed behavior can be well explained by the CEF effect with an $\sim$\,100\,K splitting
between the ground state doublet and the first excite state. The low-temperature physical properties manifest strong magnetic field dependencies.
The constructed $H-T$ phase diagram indicates that the AFM order can be suppressed by magnetic field of $H_{c}$ = 32.5\,kOe to $T$ = 0 through
MMT. No clear indication of nFL behavior was observed in the vicinity of the $H_{c}$. A FL behavior close to the $H_{c}$ is found over a limited
range of temperature and magnetic field. At higher fields an anomalous temperature dependence of the resistivity was observed.

\begin{acknowledgments}
We would like to thank Hyunjin Ko for single crystal x-ray measurements. Work at Ames Laboratory was supported by the Basic Energy Sciences, U.S.
Department of Energy under Contract No. DE-AC02-07CH11358.
\end{acknowledgments}

\clearpage

\begin{table}
\caption{\label{table1} CEF parameters, energy levels and wave functions in CeNiGe$_{3}$.}
\begin{ruledtabular}
\begin{tabular}{ccccccc}
\multicolumn{6}{l}{CEF parameters (K)} \\ \hline
 & $B^{0}_{2}$ & $B^{2}_{2}$ &  $B^{0}_{4}$ & $B^{2}_{4}$ & $B^{4}_{4}$ &  \\
 & 5.36 & -11.15 & 0.12 & 0.58 & 1.02 &  \\
  &       &       &      &      &       &             \\ \hline \hline
\multicolumn{6}{l}{Energy levels and Wave Functions}  \\ \hline $E$(K)  & $\mid +\frac{5}{2} \rangle$ & $\mid +\frac{3}{2} \rangle$ & $\mid
+\frac{1}{2}
\rangle$ & $\mid -\frac{1}{2} \rangle$ & $\mid -\frac{3}{2} \rangle$ & $\mid -\frac{5}{2} \rangle$ \\
171.7 & 0.848  & 0      & -0.357  & 0      & 0.391  & 0 \\
171.7 & 0      & -0.393 & 0       & 0.358  & 0      & -0.847 \\
99.3  & -0.527 & 0      & -0.635  & 0      & 0.565  & 0 \\
99.3  & 0      & -0.564 & 0       & 0.634  & 0      & 0.529 \\
0     & 0      & -0.726 & 0       & -0.685 & 0      & 0.047 \\
0     & 0.047  & 0      & 0.685   & 0      & 0.726  & 0 \\
\end{tabular}
\end{ruledtabular}
\end{table}

\clearpage

\end{document}